\documentclass[pra,aps,preprint,showpacs]{revtex4}
\usepackage{epsfig,amssymb,amsfonts}

\def\opone{\leavevmode\hbox{\small1\kern-3.8pt\normalsize1}}

\begin{document}

\title{Thermal entanglement and teleportation in a two-qubit Heisenberg chain with
Dzyaloshinski-Moriya anisotropic antisymmetric interaction}
\author{Guo-Feng Zhang\footnote{Corresponding author.}\footnote{Email:
gf1978zhang@buaa.edu.cn}} \affiliation{Department of physics, School
of sciences, Beijing University of Aeronautics and Astronautics,
Xueyuan Road No. 37, Beijing 100083, People's Republic of China }

\begin{abstract}
Thermal entanglement of a two-qubit Heisenberg chain in presence of
the Dzyaloshinski-Moriya (DM) anisotropic antisymmetric interaction
and entanglement teleportation when using two independent Heisenberg
chains as quantum channel are investigated. It is found that the DM
interaction can excite the entanglement and teleportation fidelity.
The output entanglement increases linearly with increasing value of
input one, its dependences on the temperature, DM interaction and
spin coupling constant are given in detail. Entanglement
teleportation will be better realized via antiferromagnetic spin
chain when the DM interaction is turned off and the temperature is
low. However, the introduction of DM interaction can cause the
ferromagnetic spin chain to be a better quantum channel for
teleportation. A minimal entanglement of the thermal state in the
model is needed to realize the entanglement teleportation regardless
of antiferromagnetic or ferromagnetic spin chains.
\end{abstract}


\maketitle

\section{Introduction}

Entanglement is one of the most fascinating features of quantum
mechanics and plays a central role in quantum information
processing. In recent years, there has been an ongoing effort to
characterize qualitatively and quantitatively the entanglement
properties of condensed matter systems and apply then in quantum
information. The quantum entanglement in solid state systems such as
spin chains is an important emerging field
\cite{man,xwa,glk,kmo,ysu,dvk,gfz,zgf}. Spin chains are natural
candidates for the realization of the entanglement compared with the
other physics systems. The Heisenberg chain, the simplest spin
chain, has been used to construct a quantum computer\cite{dlo}. By
suitable coding, the Heisenberg interaction alone can be used for
quantum computation \cite{dal,dpd,lfs}. In addition, quantum
teleportation has been extensively investigated both experimentally
and theoretically. Since the decoherence from environment always
impacts on the degree of entanglement, the resource of maximally
entangled states is hard to prepare in a real experiment. Certainly,
a mixed entangled state as the resource is approximately near to the
real circumstances. As an important source, the thermal entanglement
has been widely investigated in many previous studies. Also the
entanglement teleportation via thermal entangled states of a
two-qubit Heisenberg XX chain has been reported \cite{yey}. Yeo
\cite{yye} \emph{et al} studied the influence of anisotropy and
magnetic field on quantum teleportation via Heisenberg XY chain. But
only the spin-spin interaction was considered in those studies, the
effects of spin-orbit coupling on the entanglement and teleportation
are rarely concerned. These are the motivations of this paper.

In this paper, we investigate the influence of spin-orbit coupling
on the thermal entanglement. The information transmission by a pair
of thermal mixed states in a two-qubit Heisenberg chain in presence
of the DM anisotropic antisymmetric interaction is investigated. A
minimal entanglement in the quantum channel is needed to transfer
entanglement information. Thermal entanglement will be given in Sec.
II.  The entanglement teleportation of two-qubit pure states and its
fidelity are derived in Sec. III and IV. In Sec.V a discussion
concludes the paper.

\section{The effect of DM interaction on thermal entanglement}

In this paper, we consider the Heisenberg model with DM interaction,
which can be described by
\begin{equation}
\label{1}
H_{DM}=\frac{J}{2}[(\sigma_{1x}\sigma_{2x}+\sigma_{1y}\sigma_{2y}+\sigma_{1z}\sigma_{2z})+\overrightarrow{D}\cdot(\overrightarrow{\sigma_{1}}\times\overrightarrow{\sigma_{2}})],
\end{equation}
here $J$ is the real coupling coefficient and $\overrightarrow{D}$
is the DM vector coupling. The DM anisotropic antisymmetric
interaction arises from spin-orbit coupling \cite{idz,tmo}. The
coupling constant $J>0$ corresponds to the antiferromagnetic case
and $J<0$ to the ferromagnetic case. For simplicity, we choose
$\overrightarrow{D}=D\overrightarrow{z}$, then the Hamiltonian
$H_{DM}$ becomes
\begin{eqnarray}
\label{2}
H_{DM}&=&\frac{J}{2}[\sigma_{1x}\sigma_{2x}+\sigma_{1y}\sigma_{2y}+\sigma_{1z}\sigma_{2z}+D(\sigma_{1x}\sigma_{2y}-\sigma_{1y}\sigma_{2x})]
\nonumber
\\&=&J[(1+iD)\sigma_{1+}\sigma_{2-}+(1-iD)\sigma_{1-}\sigma_{2+}].
\end{eqnarray}
Without loose of generality, we define $|0\rangle$ $(|1\rangle)$ as
the ground (excited) state of a two-level particle. The eigenvalues
and eigenvectors of $H_{DM}$ are given by
\begin{eqnarray}
\label{3} H_{DM}|00\rangle&=&\frac{J}{2}|00\rangle,\nonumber
\\H_{DM}|11\rangle&=&\frac{J}{2}|11\rangle, \nonumber
\\H_{DM}|+\rangle&=&(J\sqrt{1+D^{2}}-\frac{J}{2})|+\rangle,\nonumber
\\H_{DM}|-\rangle&=&(-J\sqrt{1+D^{2}}-\frac{J}{2})|-\rangle,
\end{eqnarray}
where $|\pm\rangle=\frac{1}{\sqrt{2}}(|01\rangle\pm
e^{i\theta}|10\rangle)$ and $\theta=\arctan D$.

As the thermal fluctuation is introducing into the system, the state
of a typical solid-state system at thermal equilibrium (temperature
$T$) is $\rho(T)=\frac{1}{Z}e^{-\beta\emph{H}}$, where $H$ is the
Hamiltonian, $Z=tre^{-\beta\emph{H}}$ is the partition function, in
the standard basis
$\{|11\rangle,|10\rangle,|01\rangle,|00\rangle\}$, the density
matrix $\rho(T)$ can be expressed as
\begin{eqnarray}
\rho(T)=\frac{1}{Z}
\left(%
\begin{array}{cccc}
  e^{-\frac{\beta J}{2}}& 0  &0 & 0 \\
  0 & \frac{1}{2}e^{\frac{1}{2}\beta (J-\delta)}(1+e^{\beta \delta}) & \frac{1}{2}e^{i\theta}e^{\frac{1}{2}\beta (J-\delta)}(1-e^{\beta \delta}) & 0 \\
  0 &   \frac{1}{2}e^{-i\theta}e^{\frac{1}{2}\beta (J-\delta)}(1-e^{\beta \delta}) & \frac{1}{2}e^{\frac{1}{2}\beta (J-\delta)}(1+e^{\beta \delta})  & 0\\
 0 &0 & 0 & e^{-\frac{\beta J}{2}} \\
\end{array}%
\right),
\end{eqnarray}
where $Z=2e^{-\frac{\beta J}{2}}(1+e^{\beta J}\cosh\frac{\beta
\delta}{2})$, $\beta=\frac{1}{kT}$ and $\delta=2J\sqrt{1+D^{2}}$. In
the following calculation, we will write the Boltzman constant
$k=1$. The entanglement of two qubits can be measured by the
concurrence $C$ which is defined as $C=\max[0,2
\max[\lambda_{i}]-\sum^{4}_{i}\lambda_{i}]$\cite{shi}, where
$\lambda_{i}$ are the square roots of the eigenvalues of the matrix
$R=\rho S\rho^{*}S$, $\rho$ is the density matrix,
$S=\sigma_{1y}\otimes\sigma_{2y}$ and $*$ stands for the complex
conjugate. The concurrence is available, no matter whether $\rho$ is
pure or mixed. Note that we are working in units so that $D$ and $J$
are dimensionless.

Based on the definition of concurrence, we can obtain the
concurrence at the finite temperature
\begin{equation}
C_{channel}\equiv
C[\rho(T)]=\frac{2}{Z}\max\left\{\frac{1}{2}|e^{\frac{1}{2}\beta(J-\delta)}(1-e^{\beta
\delta})|-e^{-\frac{\beta J}{2}},0\right\}.
\end{equation}
The concurrence $C=0$ indicates the vanishing entanglement. The
critical temperature $T_{c}$ above which the concurrence is zero is
determined by the nonlinear equation
\begin{eqnarray}
\left\{%
\begin{array}{ll}
     e^{\frac{J}{T}}\sinh\frac{\delta}{2T}=-1, & \hbox{if $J<0$;} \\
    e^{\frac{J}{T}}\sinh\frac{\delta}{2T}=1, & \hbox{if $J>0$.} \\
\end{array}%
\right.
\end{eqnarray}
which can be solved numerically. When $D=0$, i.e. $\delta=2J$, it is
found that $T_{c}=\frac{2J}{\ln3}$ for $J>0$, but there is no
entanglement at any temperature for $J<0$. These accord to the
conclusions in Ref. \cite{xgw}. Fig.1 demonstrates the dependence of
thermal entanglement on $J$ and $D$ at $T=0.5$.
\begin{figure}
\begin{center}
\epsfig{figure=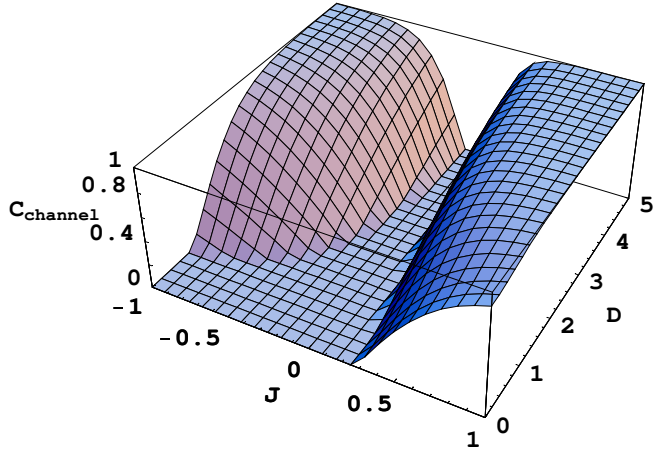}
\end{center}
\caption{(Color online) The thermal concurrence for the spin channel
when $T=0.5$. $T$ is plotted in units of the Boltzmann¡¯s constant
$k$. We work in units where $D$ and $J$ are dimensionless.}
\end{figure}
Although there is no entanglement for ferromagnetic case when $D=0$,
while $D$ increases, entanglement will be inspired and the area of
$J$ for which $C=0$ will decrease. The entanglement can reach the
maximum value by adjusting the DM interaction constant for the two
cases.

\section{Thermal entanglement teleportation}

For the entanglement teleportation of the whole two-qubit system as
the resource of the thermal mixed state in a Heisenberg spin chain,
the standard teleportation through mixed states can be regarded as a
general depolarising channel \cite{gbo,mho}. Similar to the standard
teleportation, the entanglement teleportation for the mixed channel
of an input entangled state is destroyed and its replica state
appears at the remote place after applying local measurement in the
form of linear operators. We consider as input a qubit in an
arbitrary pure state
$|\psi\rangle_{in}=\cos\frac{\theta}{2}|10\rangle+e^{i\phi}\sin\frac{\theta}{2}|01\rangle$$(0\leq\theta\leq\pi,
0\leq\phi\leq2\pi)$. Here different values of $\theta$ describe all
states with different amplitudes, and $\phi$ stands for the phase of
these states. The output state is then given by \cite{ape}
\begin{equation}
\rho_{out}=\sum_{ij}p_{ij}(\sigma_{i}\otimes\sigma_{j})\rho_{in}(\sigma_{i}\otimes\sigma_{j}),
\end{equation}
where $\sigma_{i}(i=0,x,y,z)$ signify unit matrix $I$ and three
components of Pauli matrix $\overrightarrow{\sigma}$
correspondingly, $p_{ij}=tr(E^{i}\rho(T))\cdot tr(E^{j}\rho(T))$ and
$\sum_{ij}p_{ij}=1$, $\rho_{in}=|\psi\rangle_{in}\langle\psi|$. Here
$E^{0}=|\Psi^{-}\rangle\langle\Psi^{-}|$,
$E^{1}=|\Phi^{-}\rangle\langle\Phi^{-}|$,
$E^{2}=|\Phi^{+}\rangle\langle\Phi^{+}|$,
$E^{3}=|\Psi^{+}\rangle\langle\Psi^{+}|$, in which
$|\Psi^{\pm}\rangle=(1/\sqrt{2})(|01\rangle\pm|10\rangle)$,
$|\Phi^{\pm}\rangle=(1/\sqrt{2})(|00\rangle\pm|11\rangle)$.

It follows that the concurrence of initial state $|\psi\rangle_{in}$
is $C_{in}=2|\sin\frac{\theta}{2}\cos\frac{\theta}{2}e^{i\phi}|$. We
calculate the measure of entanglement for the teleported state
$\rho_{out}$ to be
\begin{equation}
C[\rho_{out}]\equiv C_{out}=\max\left\{\frac{2\left\{C_{in}e^{\beta
J}(\sinh\frac{\beta\delta}{2})^{2}-2(1+D^{2})\cosh\frac{\beta\delta}{2}\right\}}{Z^{2}(1+D^{2})},0\right\}.
\end{equation}
From Eq.(8), it can be seen that $C_{out}$ increases linearly with
the increasing value of $C_{in}$. The result can also be seen from
Fig. 2(a) and Fig.3.
\begin{figure}
\begin{center}
\epsfig{figure=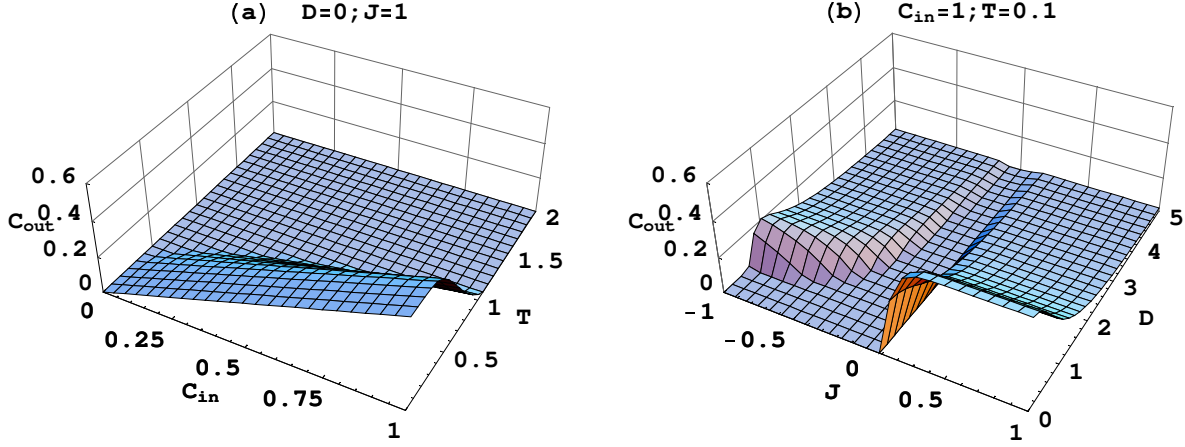}
\end{center}
\caption{(Color online) The teleported thermal concurrence $C_{out}$
as a function of the input concurrence $C_{in}$, DM interaction $D$,
spin coupling $J$ and temperature $T$. $T$ is plotted in units of
the Boltzmann¡¯s constant $k$. We work in units where $D$ and $J$
are dimensionless.}
\end{figure}
\begin{figure}
\begin{center}
\epsfig{figure=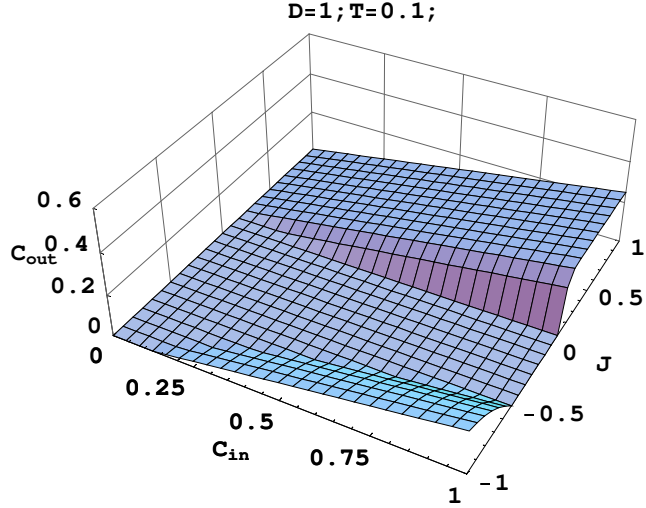}
\end{center}
\caption{(Color online) The teleported thermal concurrence $C_{out}$
as a function of the input concurrence $C_{in}$ and spin coupling
$J$ when temperature $T=0.1$ and DM interaction $D=1$. $T$ is
plotted in units of the Boltzmann¡¯s constant $k$. We work in units
where $D$ and $J$ are dimensionless.}
\end{figure}

The quantity $C_{out}$ as a function of $C_{in}$ is plotted in Fig.
2 when the DM interaction $D$, the temperature $T$ and the coupling
coefficient $J$ are changed. Fig. 2(a) is a plot of $C_{out}$ as
functions of $C_{in}$, $T$ when $D=0$ and $J=1$ for which the
critical temperature of the channel concurrence
$T_{c}=\frac{2J}{\ln3}=\frac{2}{\ln3}\approx1.82$, from Fig. 2(a),
we know that $C_{out}$ remains zero when $T>1$, so a minimal
entanglement of the thermal mixed state must be provided in such
quantum channel in order to realize entanglement teleportation. When
the initial state is in a maximum entangled state which corresponds
to Fig.2(b), $C_{out}$ exists regardless of the sign of $J$. For
$J<0$, firstly the output entanglement increases with the increasing
$D$ from zero to a certain value that is much smaller than $C_{in}$
and then begins to fall until to be zero. However, $C_{out}$
decreases monotonously with the increasing $D$ for $J>0$. It may be
advantageous for increasing $C_{out}$ and the channel entanglement
$C_{channel}$ by introducing the DM interaction for $J<0$, however,
when $J>0$ the DM interaction can only cause $C_{channel}$ increase.
As the increase of DM interaction, $C_{out}$ will decrease until to
be zero when $D$ is large for both $J>0$ and $J<0$. These are due to
the fact that $C_{out}=\max\{-4\cosh[\beta\delta/2]/Z^{2},0\}=0$
when $D\rightarrow\infty$. The maximum value of $C_{out}$ is much
smaller than that of the channel entanglement. Under the general
circumstances, the output entanglement of two-qubit state
$|\psi\rangle_{in}$ will decrease via the quantum channel. These
results can be found by comparing Fig.2 with Fig.1 \cite{pzh}.

Fig.3 shows the dependence of $C_{out}$ on $C_{in}$ and spin
coupling $J$ for a given DM interaction and temperature. As the
channel concurrence shows, $C_{out}$ behaves obviously different for
$J>0$ and $J<0$. For $J<0$, only when $|J|>0.5$, $C_{out}$ is
nonvanishing when the initial state is a maximum entangled state. If
$0<C_{in}<1$, $|J|$ must be larger in order to realize entanglement
teleportation. We can know that $C_{channel}\approx0.597$ for
$J=-0.5$ at the same condition with Fig.3, these results show again
a minimum entanglement must be provided. The same conclusion can be
obtained for $J>0$.

\section{The fidelity of entanglement teleportation}

To characterize the quality of the teleported state $\rho_{out}$, it
is often quite useful to look at the fidelity between $\rho_{out}$
and $\rho_{in}$ defined by \cite{rjo}
\begin{equation}
F(\rho_{in},\rho_{out})=\left\{tr\left[\sqrt{(\rho_{in})^{1/2}\rho_{out}(\rho_{in})^{1/2}}\right]\right\}^{2}.
\end{equation}
The concept of fidelity has been a useful indicator of the
teleportation performance of a quantum channel when the input state
is a pure state. The average fidelity $F_{A}$ of teleportation can
be formulated by
\begin{equation}
F_{A}= \frac{\int^{2\pi}_{0}d\phi\int^{\pi}_{0}F\sin\theta d
\theta}{4\pi}.
\end{equation}
\begin{figure}
\begin{center}
\epsfig{figure=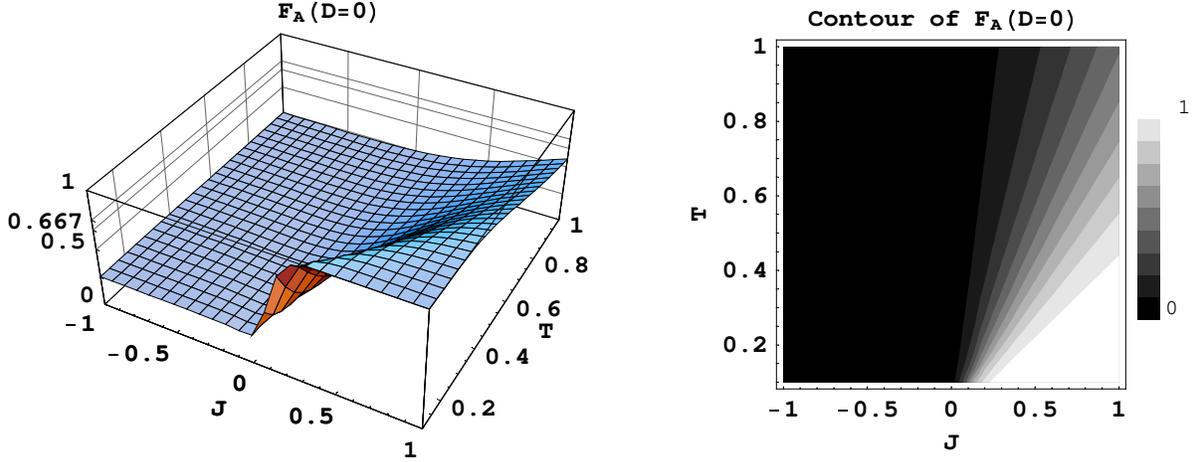}
\end{center}
\caption{(Color online) The average fidelity $F_{A}$ as a function
of spin coupling $J$ and temperature $T$ when $D=0$. $T$ is plotted
in units of the Boltzmann¡¯s constant $k$. We work in units where
$D$ and $J$ are dimensionless.}
\end{figure}
If our model is used as quantum channel, $F_{A}$ can be expressed by
\begin{equation}
F_{A}= \frac{2(1+D^{2})+e^{2\beta
J}[1+2D^{2}+(3+2D^{2})\cosh\beta\delta]}{6(1+D^{2})(1+e^{\beta
J}\cosh\frac{\beta \delta}{2})^{2}}.
\end{equation}
This is the maximal fidelity achievable from $\rho(T)$. In order to
transmit $|\psi\rangle_{in}$ with better fidelity than any classical
communication protocol, we require Eq.(11) to be strictly greater
than 2/3($\approx0.667$). When $D=0$, this requirement becomes
$e^{2\beta J}>11$.

The average fidelity $F_{A}$ is plotted as a function of spin
coupling $J$ and temperature $T$ when $D=0$ in Fig.4. When $D=0$,
$F_{A}$ is larger than $2/3$ if $0<T<\frac{2J}{\ln11}$. So $F_{A}$
is always smaller than $2/3$ for $J<0$ at any temperature. However,
for $J>0$, $F_{A}$ can arrive $1$ at near zero temperature and
begins to fall until $2/3$ at the point $T=\frac{2J}{\ln11}$. It
means that the entanglement teleportation of the mixed channel is
inferior to the classical communication when $J<0$ without DM
interaction. Fig.5 give the dependence of $F_{A}$ on DM interaction
$D$ and spin coupling $J$. By introducing the DM interaction,
$F_{A}$ can be larger than $2/3$ for $J<0$ (for example, $T=0.1$,
$J\in[-0.5,-1]$). For $J>0$, $F_{A}$ decreases monotonously with the
increasing $D$. When the DM interaction is very strong, $F_{A}$
approaches infinitely the value of 2/3 for both the two cases. These
results show that we must strengthen the DM interaction to be a
certain value in order to use ferromagnetic spin chain as a quantum
channel for entanglement teleportation, which is contrary to
antiferromagnetic spin chain.
\begin{figure}
\begin{center}
\epsfig{figure=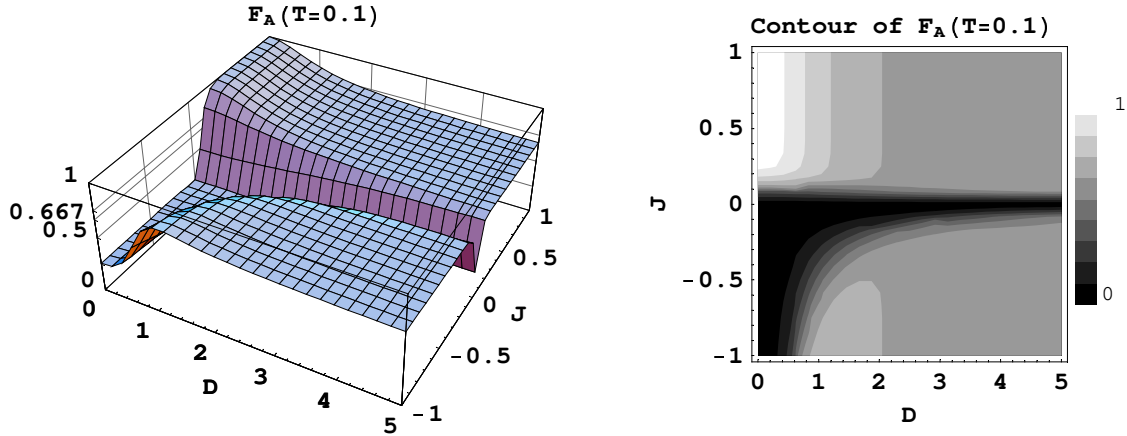}
\end{center}
\caption{(Color online) The average fidelity $F_{A}$ as a function
of $J$ and $D$ for a given temperature. $T$ is plotted in units of
the Boltzmann¡¯s constant $k$. We work in units where $D$ and $J$
are dimensionless.}
\end{figure}
\section{Conclusions}
We have investigated the thermal entanglement of two-qubit spin
chain with DM anisotropic antisymmetric interaction and entanglement
teleportation via the model. The entanglement can reach the maximum
value by adjusting the DM interaction constant for ferromagnetic and
antiferromagnetic case. By introducing the DM interaction, the
output entanglement and fidelity can be increased for ferromagnetic
case, which are contrary to antiferromagnetic case. When the DM
interaction is very strong, the average fidelity of entanglement
teleportation will approach a fixed value that is the maximal one
for classical communication. A minimal entanglement of the thermal
state in the model is needed to realize the entanglement
teleportation.

\section{acknowledgements}
This work was supported by the National Natural Science Foundation
of China (Grant No. 10604053).


\begin{references}
\bibitem{pzh} In order to see that the effect of DM interaction on the entanglement, we take $T=0.5$ in Fig.1.
                Although the temperature is different in Fig.1 and
                Fig.2, we can also compare the entanglement between Fig.1 and Fig.2. since the temperature plays a negative
                effect.
\end{references}
\end{document}